\documentclass[twocolumn,showpacs,preprintnumbers,amsmath,amssymb,floatfix]{revtex4-1}
\usepackage{graphicx}
\usepackage{epstopdf}
\usepackage{afterpage}
\usepackage{amsmath}
\usepackage{dcolumn}
\usepackage{bm}

\begin{document}

\preprint{APS/123-QED}

\title{Conversion of bright magneto-optical resonances into dark at fixed laser frequency 
for $D_2$ excitation of atomic rubidium}

\author{M.~Auzinsh}%
\email{Marcis.Auzins@lu.lv}%
\author{A.~Berzins}%
\author{R.~Ferber}%
\author{F.~Gahbauer}%
\author{L.~Kalvans}%
\author{A.~Mozers}%
\author{D.~Opalevs}%

\affiliation{Laser Centre, The University of Latvia, 19 Rainis
Boulevard, LV-1586 Riga, Latvia}%


\date{\today}

\begin{abstract}
Nonlinear magneto-optical resonances on the hyperfine transitions belonging to the $D_2$ line of rubidium were changed 
from bright to dark resonances by changing the laser power density of the single exciting laser field or by changing 
the vapor temperature in the cell.  
In one set of experiments atoms were excited by
linearly polarized light from an extended cavity diode laser with polarization vector perpendicular to the 
light's propagation direction and magnetic field, and laser induced fluorescence (LIF) was observed along the direction 
of the magnetic field, which was scanned. A low-contrast bright resonance was observed 
at low laser power densities when the laser was tuned to the $F_g=2 \longrightarrow F_e=3$ transition of $^{87}$Rb 
and near to the $F_g=3 \longrightarrow F_e=4$ transition of $^{85}$Rb. The bright 
resonance became dark as the laser power density was increased above 0.6mW/cm$^2$ or 0.8 mW/cm$^2$, respectively. 
When the $F_g=2\longrightarrow F_e=3$ transition of $^{87}$Rb was excited with circularly polarized light in 
a second set of experiments, a bright 
resonance was observed, which became dark when the temperature was increased to around 50$^o$C. 
The experimental observations at room temperature could be reproduced with 
good agreement by calculations based on a theoretical model, although the theoretical model was not able to 
describe measurements at elevated temperatures, where reabsorption was thought to play a decisive role. 
The model was derived from the optical Bloch equations and 
included all nearby hyperfine components, averaging over the Doppler profile, mixing of magnetic sublevels in the 
external magnetic field, and a treatment of the coherence properties of the exciting radiation field. 
\end{abstract}
\maketitle

\section{\label{Intro:level1}Introduction}  
  
  When laser fields create a $\Lambda$-system, coherences between the two lower states can be manipulated by magnetic fields 
to give rise to resonances. When two separate radiation fields act as the "legs" of the $\Lambda$-scheme, one speaks of 
electromagnetically induced transparency (EIT)~\cite{Harris:1990,Fleischauer:2005} or electromagnetically induced 
absorption (EIA)~\cite{Akulshin:1998}. However, the $\Lambda$-system can also be created by a single radiation field when 
the two lower levels are different magnetic sublevels of the same state. In this case, one often refers to the phenomenon 
as a nonlinear magneto-optical resonance, which can be bright or dark, as the absorption at zero magnetic field represents 
a local maximum or minimum, respectively. 
Recently, attention has been focused on the conversion of EIT to EIA as a function of the properties of the pump field, 
such as  polarization~\cite{Brazhnikov:2010} and laser power density~\cite{Kim:2005,Dahl:2008,Zhukov:2009,Ram:2010}.
In this work we describe the conversion of a bright nonlinear magneto-optical resonance to a dark resonance as a function 
of the laser power density of the single optical field and the temperature. 
A change from bright to dark with increasing laser power density had been 
observed before in the cesium $D_1$ transition~\cite{Papoyan:2002}, but could not be reproduced by the theoretical model. 
Another measurement for the rubidium $D_2$ transition found that the contrast of a bright resonance peaks rapidly at low 
laser power densities and then decreases with increasing laser power density, but never becomes a dark 
resonance~\cite{Mijailovic:2007}. Elsewhere the temperature dependence of the contrasts of bright and dark 
resonances in the $D_1$ transition of sodium was investigated, but no change from bright to dark resonance was 
observed\cite{Alzetta:2003}. 
In the present study the change from bright to dark resonance 
as a function of laser power density was observed and could be described with good agreement by a 
theoretical model based on the optical Bloch equations. 
In addition, for circularly polarized exciting laser radiation, a bright resonance at room temperature was observed to 
become dark at elevated temperatures.  

  Dark, nonlinear magneto-optical resonances were first observed in cadmium vapor by Lehmann and Cohen-Tannoudji 
in 1964~\cite{Lehmann:1964}.  
Much later in 2000, Dancheva \emph{et al.} observed bright resonances~\cite{Dancheva:2000}, 
in which the laser induced fluorescence (LIF) as a function of 
magnetic field has a maximum at zero magnetic field. It was shown based on theoretical considerations that 
dark resonances should be expected for linearly polarized exciting radiation when the total angular momentum of the 
ground state $F_g$ is greater than or equal to the total angular momentum of the excited state $F_e$ ($F_g \geq F_e$), 
while a bright resonance should be expected when $F_e > F_g$~\cite{Renzoni:2001a,Alnis:2001,Papoyan:2002}. 
In order to calculate the signals from nonlinear magneto-optical resonances, models based on the optical Bloch 
equations were developed. In 1978, Picqu\'e used a model based on the optical Bloch equations to describe 
dark resonances in a beam of sodium atoms at $D_1$ excitation~\cite{Picque:1978}. 
In 2002 Andreeva \emph{et al.} investigated nonlinear experimental magneto-optical resonances using 
a model based on the optical Bloch equations that took into account all neighboring hyperfine transitions and 
averaged over the Doppler profile and laser beam transit time distribution~\cite{Andreeva:2002}.  
Around the same time, the $F_g=4\longrightarrow F_e=3,4,5$ transition of cesium was observed to be bright when 
the laser power density of the exciting radiation was 30 mW/cm$^2$, but dark at a laser power density of 
600 mW/cm$^2$. In that case, the theoretical model was not able to reproduce this change, possibly because it 
was limited to the cycling transitions~\cite{Papoyan:2002}.
In time, detailed frameworks
suitable for numerical integration were developed (see, for example,~\cite{Malakyan:2004,Blushs:2004}).
Detailed agreement between theoretical and experimental signals for nonlinear magneto-optical resonances were 
presented, for example, in~\cite{Auzinsh:2008, Auzinsh:2009}.

  Although basic theoretical considerations suggest that for linearly polarized exciting radiation,  
resonances should be bright when $F_e > F_g$ and dark when $F_g \geq F_e$  
nevertheless the shape of real resonances depends also on other 
experimental parameters. Working with two coherent laser fields, Kim \emph{et al.} observed the conversion of 
EIT to EIA as a function of the coupling laser power in the $D_1$ line of rubidium~\cite{Kim:2005}. Other groups 
have also studied the influence of coupling laser power on EIT and EIA in different systems and for different 
polarizations of the coherent fields~\cite{Dahl:2008, Zhukov:2009, Zigdon:2009, Ram:2010}. 
Brazhnikov \emph{et al.} have studied the effect of the polarization of counterpropagating fields on 
EIT and EIA~\cite{Brazhnikov:2010}. Even a small change in the ellipticity of one of the fields can change EIT to EIA. 
In the case of nonlinear magneto-optical resonances with a single laser field, it was observed that a dark resonance 
can become bright when the laser frequency is changed by only a few hundred MHz~\cite{Auzinsh:2009}. 

  In this work, we report that when a single, linearly polarized laser field was tuned to 
the $F_g=2 \longrightarrow F_e=3$ transition of $^{87}$Rb and the $F_g=3 \longrightarrow F_e=4$ transition of $^{85}$Rb,
bright resonances became dark when the laser power density was increased. In the case of circularly polarized excitation, 
the bright resonance became dark at higher temperatures.  
For sufficiently small temperatures, where reabsorption effects were not significant, 
experimental measurements were reproduced by theoretical calculations. The sensitivity of the resonances' shapes to 
the experimental parameters provided a stringent test of the theoretical model.

\begin{figure}[htbp]
	\centering
		\resizebox{\columnwidth}{!}{\includegraphics{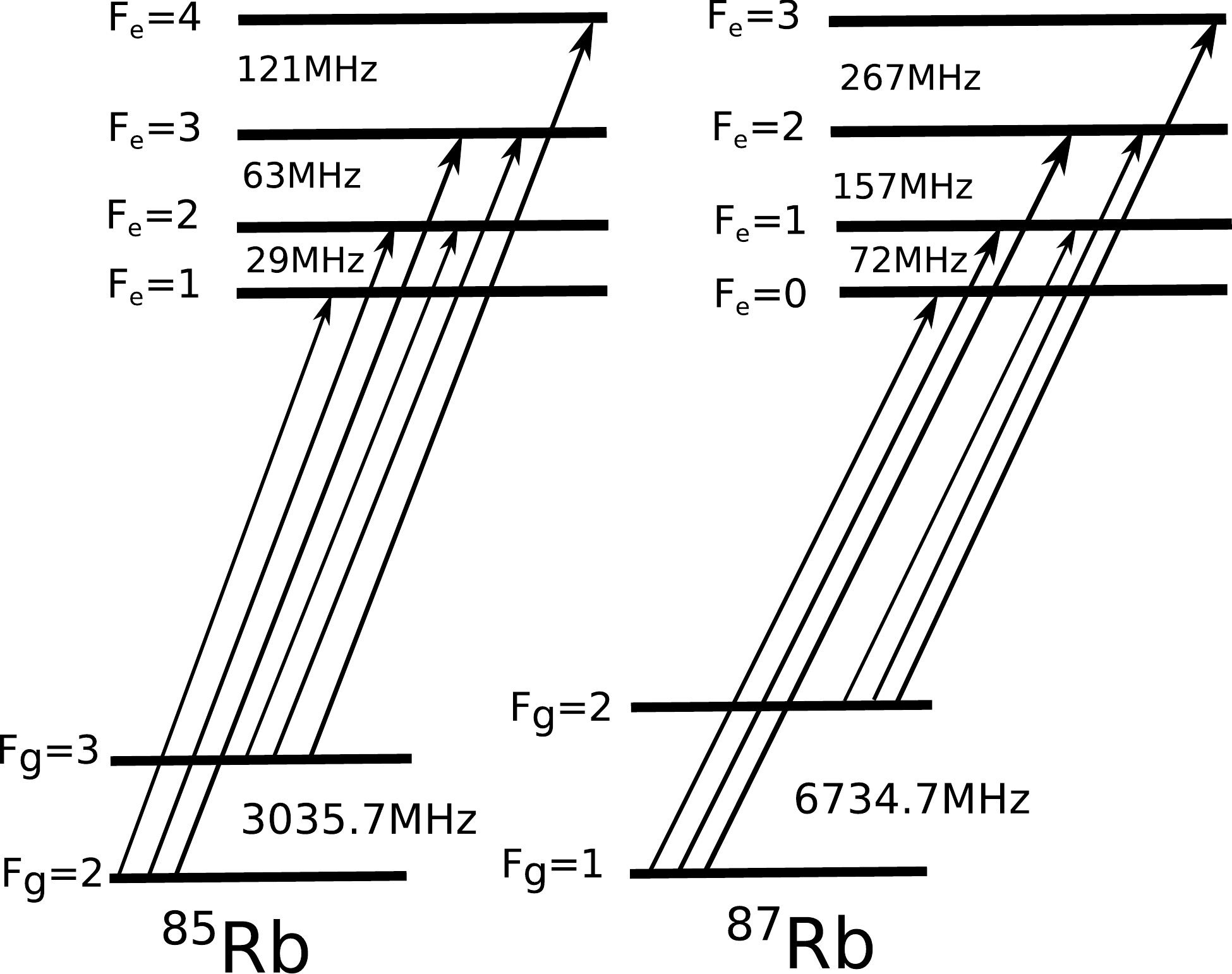}}
	\caption{\label{fig:levels} Hyperfine level structure and transitions of the  $D_2$ line of rubidium. }
\end{figure}

\section{\label{Experiment:level1}Experimental Description}
Figure~\ref{fig:levels} shows the level scheme of the $D_2$ line of rubidium. Note that the $^{85}$Rb 
hyperfine transitions are essentially unresolved under the typical Doppler broadening at room temperature, which has a  
full width at half maximum (FWHM) of
 approximately 500 MHz, while the $^{87}$Rb hyperfine transitions are partially 
resolved.

\begin{figure}[htbp]
	\centering
		\resizebox{0.5\columnwidth}{!}{\includegraphics{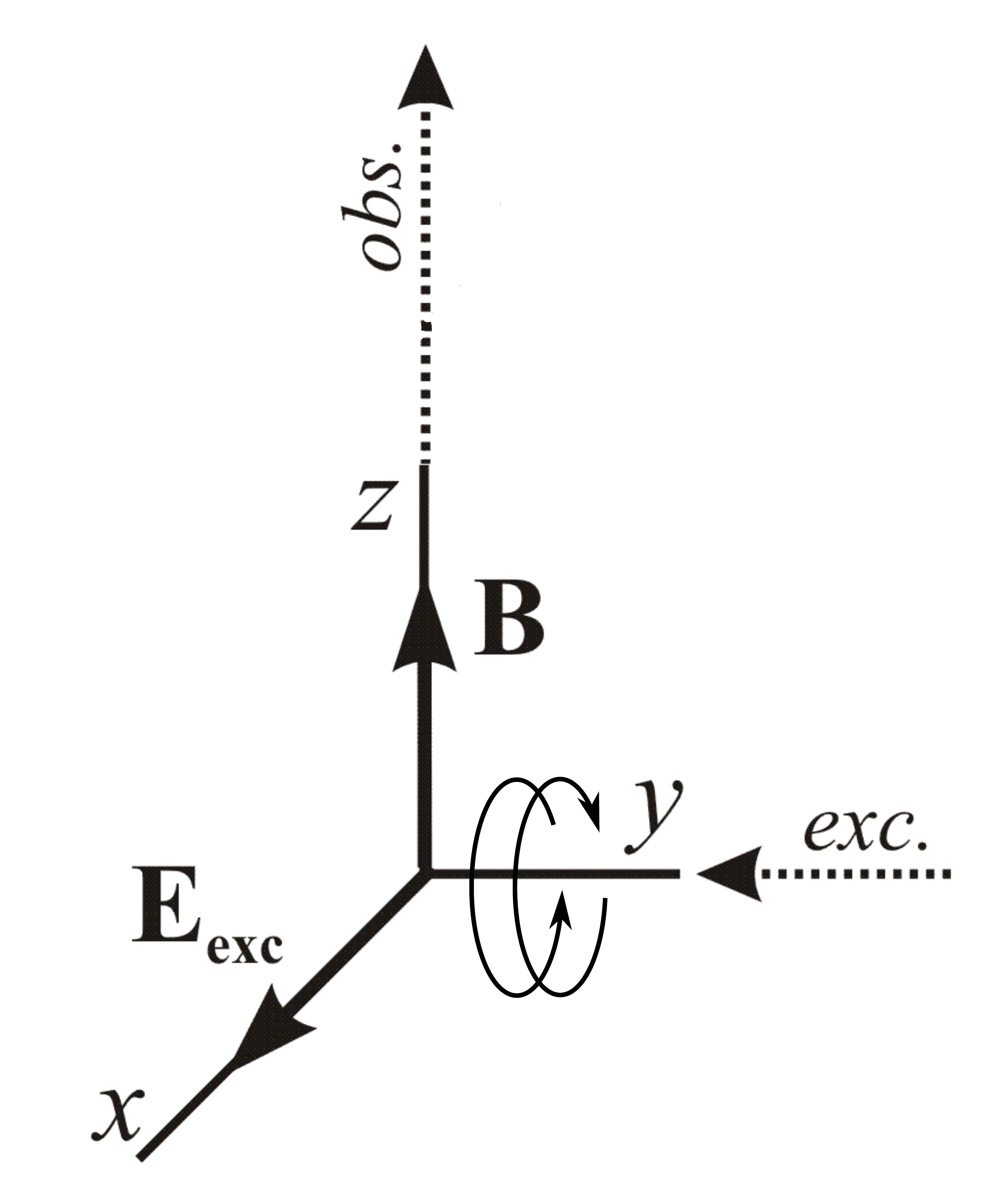}}
	\caption{\label{fig:geometry} Experimental geometry. The relative orientation of the laser beam (\textit{exc}), 
laser light polarization (\textbf{E$_{exc}$}), magnetic field (\textbf{B}), and observation direction (\textit{obs}). 
Circularly polarized light could be produced by means of a $\lambda/4$-plate. 
}
\end{figure}

The geometry of the excitation in the experiment is given in Fig.~\ref{fig:geometry}. The exciting laser 
radiation was linearly polarized with polarization vector perpendicular to the magnetic field and to the 
propagation direction of the radiation. For some experiments, a Thorlabs achromatic $\lambda/4$ plate was inserted 
after the linear polarizer to produce circularly polarized radiation in the cell. 
The observation direction was parallel to the magnetic field. 

The laser was an external cavity diode laser. The temperature of the box and the diode were stabilized by 
Thorlabs TED200 temperature controllers, and the current was controlled by a Thorlabs LDC205B current controller. 
The laser frequency was stabilized manually, and the frequency drift during a typical data-taking run was typically 
around 10 MHz or less. 
The wavelength was monitored by a HighFinesse WS-7 wavemeter. 
A Thorlabs BP104-VIS beam profiler was used to measure the beam diameter, which was taken to be the full width at 
half maximum of the intensity distribution. The cell was a 25 mm long pyrex cell with optical quality windows filled 
with natural rubidium and produced by Toptica, A.G. The laser radiation entered the cell through a Glan-Thompson 
polarizer. A combination of neutral density filters and a polarization rotator before the polarizer was used to 
select various laser power densities. 
The laser induced fluorescence (LIF) was focused onto a Thorlabs FDS100 photodiode and 
amplified. Signals were recorded by an Agilent DSO5014A oscilloscope. 

The cell was located at the center of a three-axis Helmholtz coil. Two coils were used to compensate the ambient 
magnetic field. The residual magnetic field was estimated to be less than 10 mG. 
The third coil was used to scan the magnetic field along the $z$-axis as indicated in 
Fig.~\ref{fig:geometry} using a bipolar BOP-50-8-M power supply from Kepco. 
In order to find the appropriate current for the field compensation, the polarization vector of the laser beam was rotated 
parallel to the $z$-axis and the magnetic field was scanned along the $z$-axis while the current in the 
compensating coils was adjusted in order to eliminate the magneto-optical resonance that appears 
when the field is not compensated.
The cell was at room temperature, except for those experiments that investigated the temperature dependence, 
when the cell was heated with hot air. The air was heated some 
distance away from the cell to avoid stray magnetic fields from the heating currents. The temperature was measured with 
two thermocouples near the cell, though the thermocouples were removed during the measurements as they were slightly 
magnetized.

\section{\label{Theory:level1}Theoretical Model}
The atomic state was described by a density matrix $\rho$ in the $\xi,F,m_F$ basis, with $F$ denoting the total atomic 
angular momentum quantum number (that characterizes the hyperfine structure), $m_F$, the respective magnetic quantum number 
for Zeeman splitting and $\xi$, the combination of quantum numbers that are the same for the transition's 
ground ($5S_{1/2}$) or excited states ($5P_{3/2}$). 
The fluorescence signal can be calculated from the part of the density matrix 
that describes the population and Zeeman coherences of the excited state $\rho_{e_ie_j}$. 
To do so, the optical Bloch equations (OBEs) that describe the time evolution of a density matrix 
were employed~\cite{Stenholm:2005}:
\begin{equation}
	i\hbar\frac{\partial\rho}{\partial t}=\left[\hat H,\rho\right]+i\hbar\hat R\rho.
	\label{eq:obe}
\end{equation}
We assumed that the interaction between the atoms and the exciting field was described in the dipole approximation by
the interaction operator $\hat V=-\mathbf{\hat d}\cdot \mathbf{E}(t)$. 
The electric field $\mathbf{E}(t)$ was treated as a classically oscillating field with a stochastically fluctuating phase. 
As the energy shifts due to the magnetic field were small compared to the fine-structure splitting, 
the Hamiltonian of the atomic interaction with the magnetic field could be written as
\begin{equation}
	\hat H_B = \frac{\mu_B}{\hbar}\left(g_J\mathbf{J}+g_I\mathbf{I}\right)\cdot\mathbf{B},
	\label{eq:b-ham}
\end{equation}
and expanded as shown, for example, in~\cite{Steck:rubidium85}. 
We denote the unperturbed atomic Hamiltonian as $\hat H_0$, and so the full Hamiltonian of the system is
\begin{equation}
	\hat H = \hat H_0 + \hat H_B + \hat V.
	\label{eq:full-ham}
\end{equation}
The relaxation operator $\hat R$ in \eqref{eq:obe} consists of the spontaneous emission rate, 
equal to the natural linewidth $\Gamma$ of the transition and the transit relaxation rate $\gamma$, which comes about 
because the moving atoms spend a finite time in the laser beam. 
The rate of atom-atom collisions in our experimental conditions (vacuum cell at room temperature) were estimated 
to be at least two orders of magnitude less than the transit relaxation rate and therefore were neglected.

The OBEs \eqref{eq:obe} were treated to obtain rate equations for the Zeeman coherences of the ground and excited states. 
The treatment included the rotating wave approximation~\cite{Allen:1975}, averaging and decorrelation of 
the stochastic phase~\cite{Kampen:1976}, and adiabatic elimination of the optical coherences 
(see~\cite{Blushs:2004} for details). 
As a result the following rate equations were obtained:
\begin{subequations} 
\label{eq:zc}
\allowdisplaybreaks 
\begin{align}
\frac{\partial \rho_{g_ig_j}}{\partial t} =& \left(\Xi_{g_ie_m} + \Xi_{e_kg_j}^{\ast}\right)\sum_{e_k,e_m}d_{g_ie_k}^\ast d_{e_mg_j}\rho_{e_ke_m} -\nonumber\\ 
-&\sum_{e_k,g_m}\Big(\Xi_{e_kg_j}^{\ast}d_{g_ie_k}^\ast d_{e_kg_m}\rho_{g_mg_j} + \nonumber \\ 
+&\Xi_{g_ie_k}d_{g_me_k}^\ast d_{e_kg_j}\rho_{g_ig_m}\Big) - \nonumber\\
-& \dot\imath\omega\rho_{g_ig_j} - \gamma\rho_{g_ig_j} + \sum_{e_ke_l}\Gamma_{g_ig_j}^{e_ke_l}\rho_{e_ke_l} + \lambda\delta(g_i,g_j) \label{eq:zcgg} \\
\frac{\partial \rho_{e_ie_j}}{\partial t} =& \left(\Xi_{e_ig_m}^\ast + \Xi_{g_ke_j}\right)\sum_{g_k,g_m}d_{e_ig_k} d_{g_me_j}^\ast\rho_{g_kg_m} -\nonumber\\
-& \sum_{g_k,e_m}\Big(\Xi_{g_ke_j}d_{e_ig_k} d_{g_ke_m}^\ast\rho_{e_me_j}+ \nonumber \\
+& \Xi_{e_ig_k}^\ast d_{e_mg_k} d_{g_ke_j}^\ast\rho_{e_ie_m}\Big) -\nonumber \\
-& \dot\imath\omega\rho_{e_ie_j} - (\Gamma + \gamma)\rho_{e_ie_j}. \label{eq:zcee}
\end{align}
\end{subequations}
In \eqref{eq:zc} the Zeeman coherences of the ground and excited states are denoted by $\rho_{g_ig_j}$ 
and $\rho_{e_ie_j}$, respectively. Each term of the rate equations describes a well defined part of the atom-light 
interaction process. Thus, the first terms in both \eqref{eq:zcgg} and \eqref{eq:zcee} describe the population 
increase and the creation of Zeeman coherences, and the second terms denote the population decrease and the destruction 
of Zeeman coherences in the respective atomic state (ground or excited) 
due to the laser-induced transitions; $d_{e_ig_j}$ is the dipole transition matrix element between the 
ground state $i$ and the excited state $j$~\cite{Auzinsh:2009b}, while $\Xi_{e_ig_j}$ describes the rate of the 
atomic transitions induced by the exciting radiation and is defined below \eqref{eq:pumping}. 
The third term of equations \eqref{eq:zc} describes the destruction of the Zeeman coherences by the 
external magnetic field; $\omega_{ij}$ is the splitting of the Zeeman sublevels. 
The fourth term describes the relaxation processes, and the fifth and sixth (absent in \eqref{eq:zcee}) 
describe the repopulation of the ground state due to spontaneous transitions and "fresh" atoms flying into 
the region of the interaction. It was assumed that the atomic equilibrium density outside the interaction region 
is normalized to unity, and so $\lambda = \gamma$. 
The transit relaxation rate was used as an adjustable model parameter and corresponds to the average time spent by 
the atoms in the interaction region.

The interaction strength $\Xi_{e_ig_j}$ is given by
\begin{equation}\label{eq:pumping}
\Xi_{e_ig_j} = \frac{\vert\varepsilon_{\bar\omega}\vert^2}{\hbar^2}\frac{1}{\frac{\Gamma+\gamma+\Delta\omega}{2}+\dot\imath\left(\bar\omega-\mathbf{k}_{\bar\omega}\mathbf{v}-\omega_{e_ig_j}\right)},
\end{equation}
 $\Delta\omega$ is the linewidth and $\bar\omega$ is the central frequency of the exciting radiation, 
$\mathbf{k}_{\bar\omega}\mathbf{v}$ is the frequency shift due to the Doppler effect. 
The value $\frac{\vert\varepsilon_{\bar\omega}\vert}{\hbar}$ is proportional to the Rabi frequency, 
which describes the coupling strength induced by the oscillating electric field between the chosen atomic states:
\begin{equation}\label{eq:Rabi}
	\Omega_R^2=\frac{\vert\varepsilon_{\bar\omega}\vert^2}{\hbar^2}\Vert d\Vert^2,
\end{equation}
where $\Vert d\Vert$ is the reduced dipole transition matrix element, which is equal for all $d_{ij}$ 
elements in Eqs. (\ref{eq:zc}). The Rabi frequency squared is proportional to the exciting radiation's power 
density with some proportionality coefficient $k_{Rabi}$,
\begin{equation}\label{eq:kRabi}
	I=k_{Rabi}\Omega_R^2,
\end{equation}
and is used as an adjustable parameter that corresponds to the exciting radiation's power density.

The experiments took place under stationary excitation conditions, for which the equations \eqref{eq:zc} are valid.  
Thus $\frac{\partial \rho_{g_ig_j}}{\partial t} = \frac{\partial \rho_{e_ie_j}}{\partial t} = 0$ 
and the differential equations \eqref{eq:zc} were reduced to a system of linear equations that could be solved 
to obtain the density matrix components that describe the Zeeman coherences of both the atomic ground and excited states. 
From the obtained density matrix the fluorescence signal was calculated as
\begin{equation} \label{fluorescence}
I_{fl}(\vec e) = \tilde I_0 \sum_{g_i,e_j,e_k} d_{g_ie_j}^{(ob)\ast}d_{e_kg_i}^{(ob)}\rho_{e_je_k},
\end{equation}
where $\tilde I_0$ is a constant of proportionality.

A summation over the different atomic velocity groups, associated to different $\mathbf{v}$ in~\eqref{eq:pumping},  
was performed to describe the Doppler broadening of the 
transitions. When the density matrix and the fluorescence signal were calculated, we took into account 
all nearby hyperfine transitions of both the ground and the excited state, 
represented by different $w_{e_ig_j}$ in~\eqref{eq:pumping}. 
Atomic constants were taken from~\cite{Steck:rubidium85,Steck:rubidium87}.
Some parameters could not be determined directly with certainty. Starting from reasonable estimates, these parameter 
values were varied in different sets of simulations in order to find the set of parameters that provided the best overall 
agreement between experiment and theory. We required the parameter values to be consistent for all measurements even if 
the agreement between theory and experiment for individual cases was not the best that could have been achieved by tailoring 
the parameters for each case. 
Thus, the laser linewidth was assumed to be $\Delta \omega=2$ MHz.
The proportionality constant $k_{Rabi}$ between the laser power density $I$ and the square of the Rabi frequency $\Omega$ 
($I=k_{Rabi}\Omega^2$) was $k_{Rabi}=0.575$ with the Rabi frequency given in MHz and the laser power density in 
mW/cm$^2$. 
The relationship between the transit relaxation rate $\gamma$ and the laser beam diameter $d$ was $\gamma=0.0033/d$ with 
the transit relaxation rate in MHz and the laser beam diameter in cm. 
The standard deviation of the Doppler profile was assumed to be 216 MHz and it was sampled with a step size of less than 2 MHz during 
the averaging over the Doppler distribution. 
No additional background was subtracted beyond the background determined during the experiment by blocking the laser beam.

\section{\label{Results:level1}Results and Discussion}
 The fluorescence signals recorded as a function of magnetic field usually showed two kinds of structure. One structure was 
broad, with a width on the order of several Gauss. This structure was caused by coherences in the excited state as well 
as by Zeeman sublevels being shifted out of resonance with the laser. The broad structure also tended to have 
a strong contrast, on the order of several percent or even tens of percent. In addition to the broad structure, 
it was also possible to observe narrow features, 
centered around zero magnetic field, with widths on the order of several hundred milligauss. These resonances are 
related to the destruction of coherences in the ground state as a magnetic field breaks the degeneracy of the magnetic 
sublevels. The features were also very small, with contrasts from a fraction of a percent to a few percent. 
(We defined the contrast of the narrow structure 
with respect to the estimated zero-field value of the curve corresponding to the larger 
structure.) Our study was focused on these narrow resonances 
and the way their contrast and even their sign depended on the laser power density and temperature. Thus, in the results 
that follow, only a narrow range of magnetic field is shown. 

\begin{figure*}[htbp]
	\centering
  \begin{tabular}{ccc}
 		\resizebox{6.0cm}{!}{\includegraphics{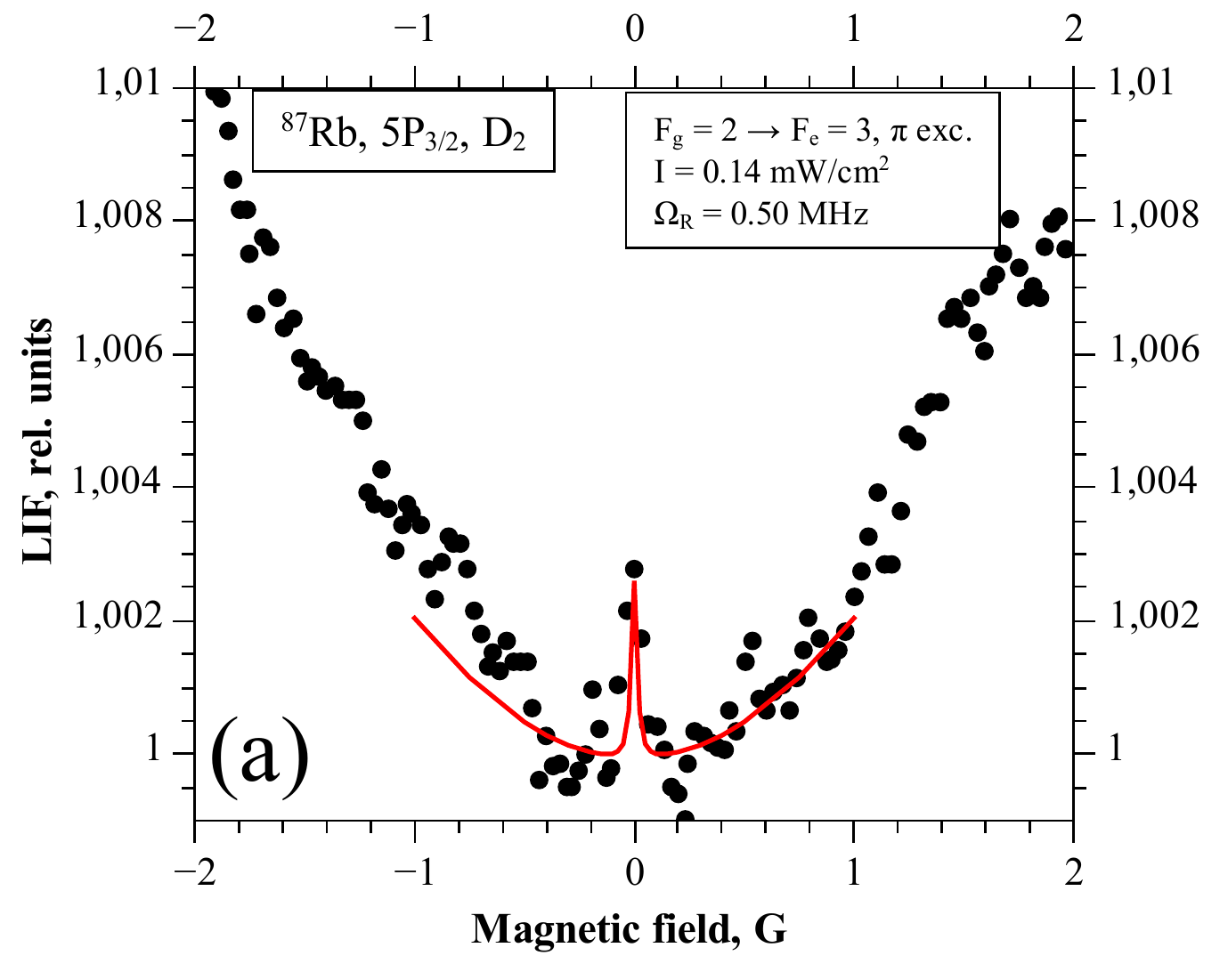}}&
		\resizebox{6.0cm}{!}{\includegraphics{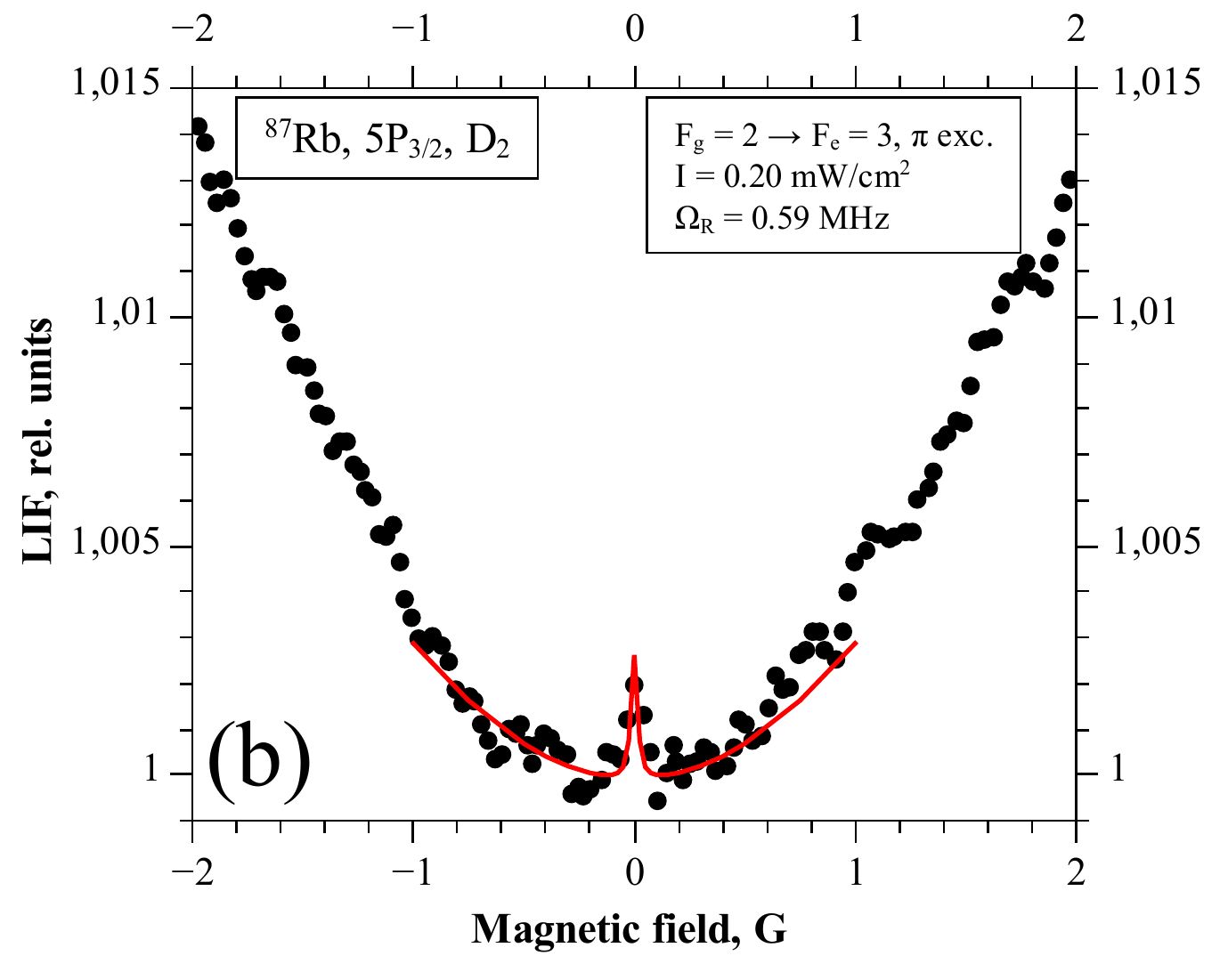}}&
		\resizebox{6.0cm}{!}{\includegraphics{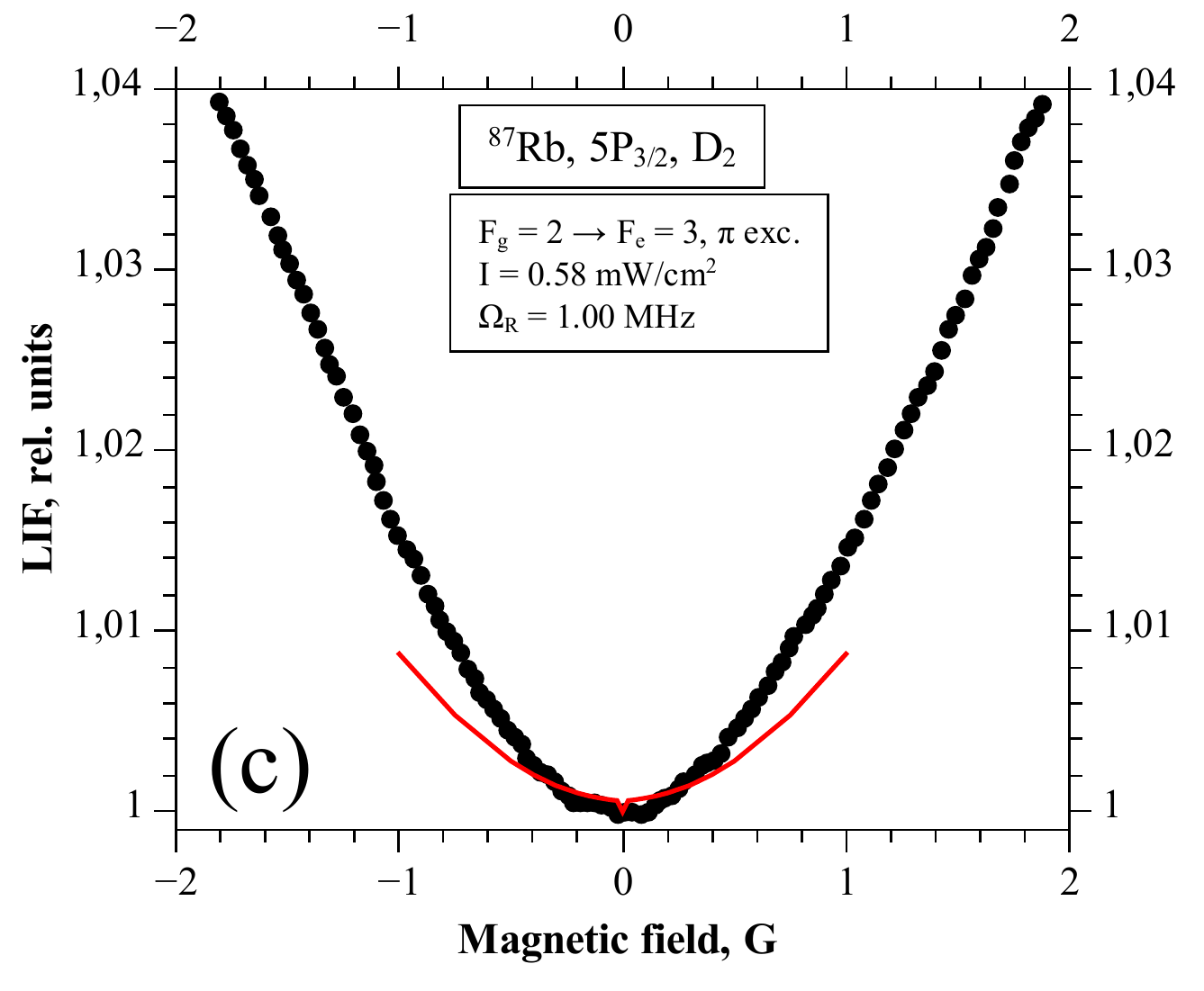}}\\
		\resizebox{6.0cm}{!}{\includegraphics{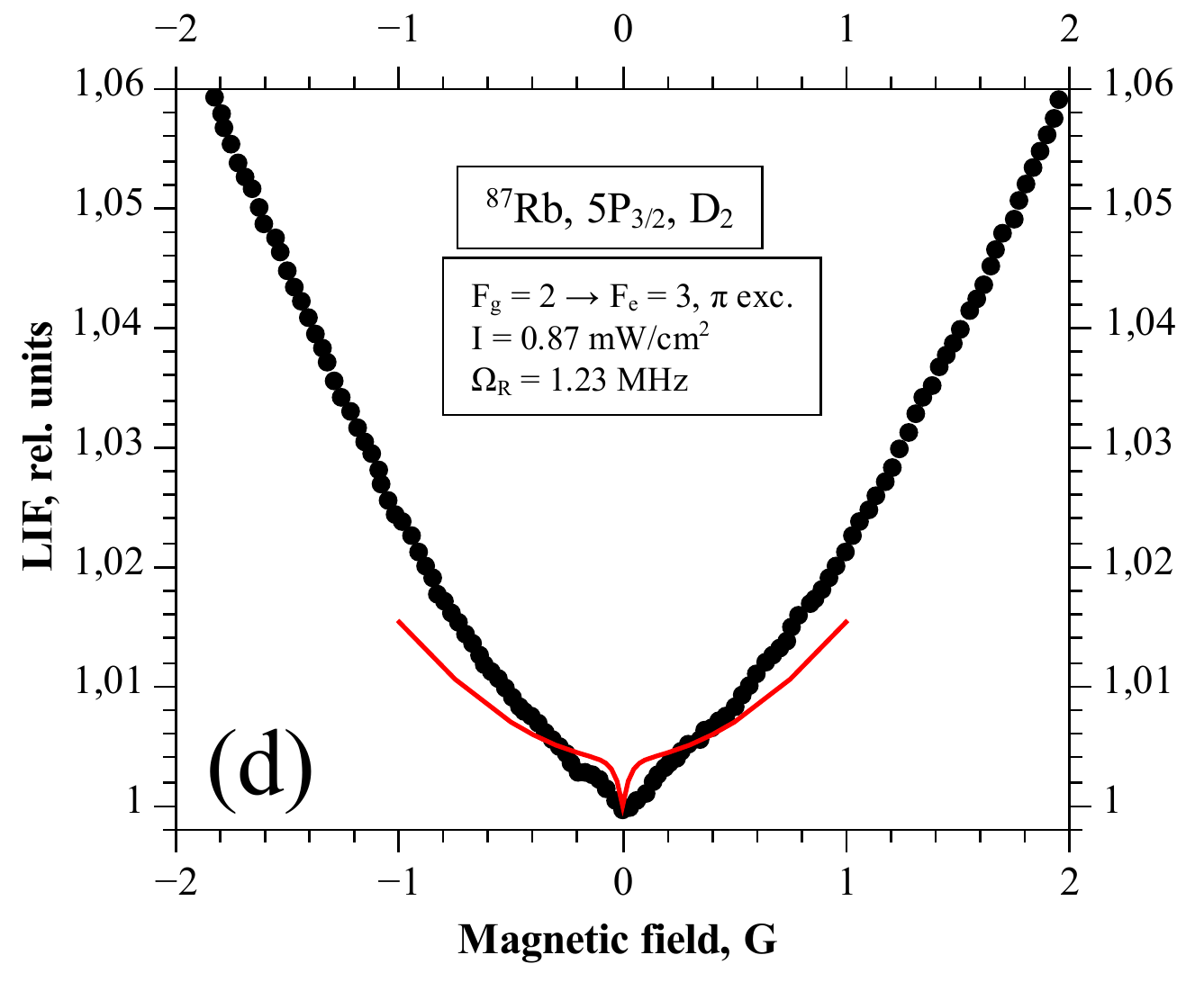}}&
		\resizebox{6.0cm}{!}{\includegraphics{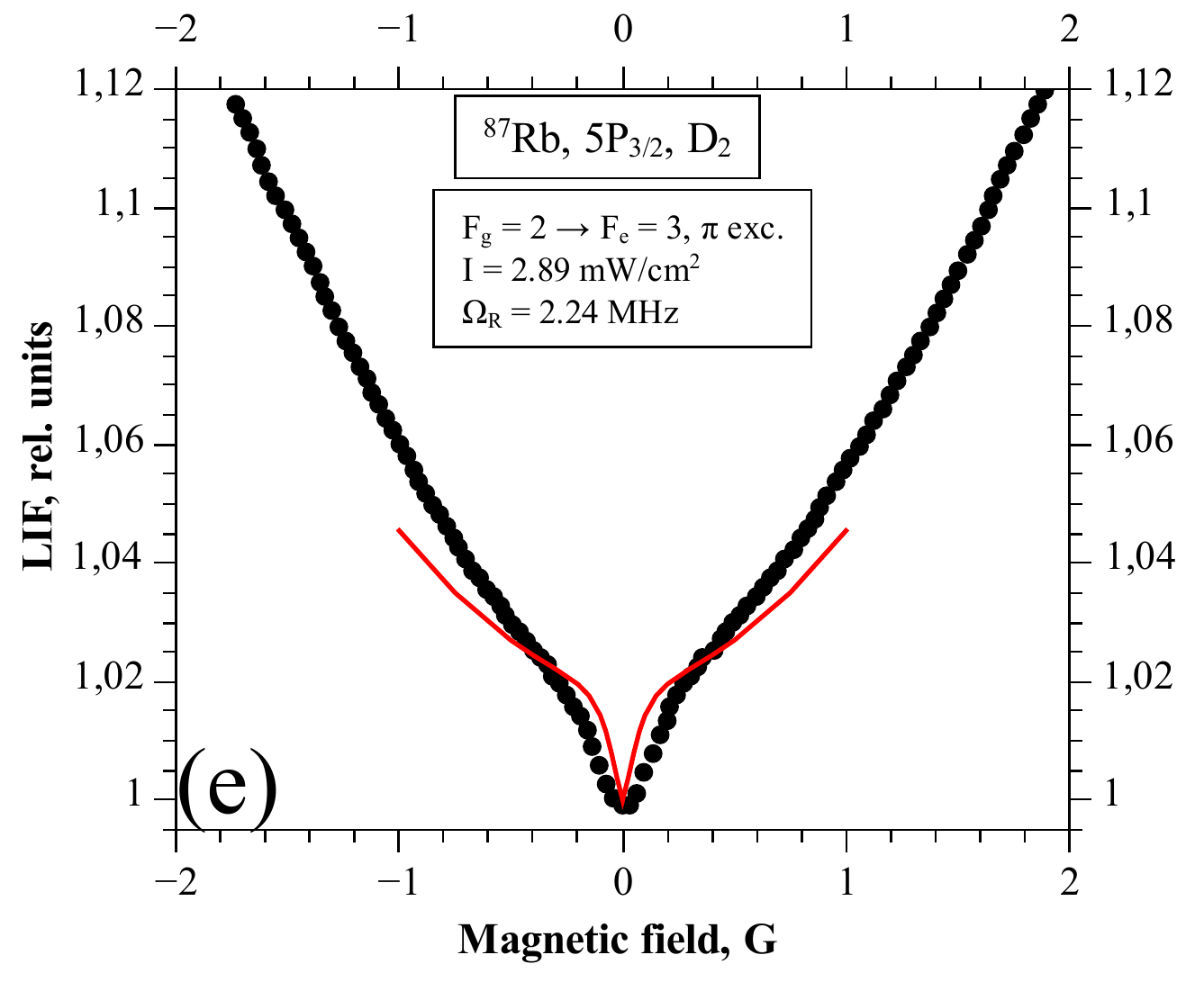}}&
		\resizebox{6.0cm}{!}{\includegraphics{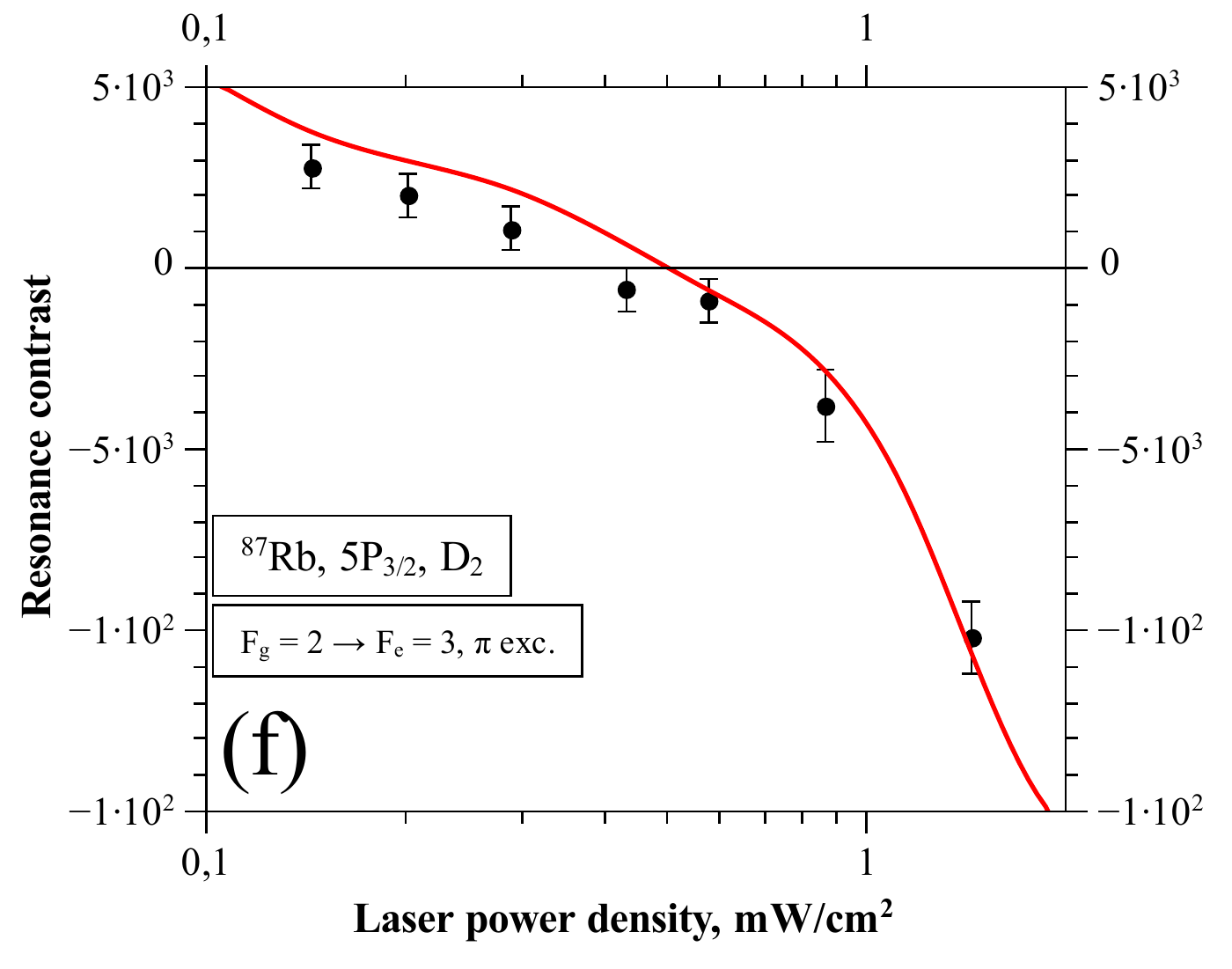}}   
   \end{tabular}
	\caption{\label{fig:rb87-23} Bright and dark resonances for various laser power densities with the 
laser tuned to the  $F_g=2\longrightarrow F_e=3$ transition of $^{87}$Rb for linearly polarized exciting laser 
radiation with a beam diameter of 
2.1 mm. Markers represent the experimental results, whereas the solid line represents the results of 
calculations. The final panel shows the contrast of the narrow resonance referenced to the estimated 
background due to the broad structure at zero magnetic field.  
}
\end{figure*}

Figure~\ref{fig:rb87-23} shows the LIF intensity as a function of magnetic field for various values of the 
power density of the exciting laser radiation with the laser tuned to the 
$F_g=2\longrightarrow F_e=3$ transition of $^{87}$Rb. The laser beam diameter was 2.1 mm. 
This transition could be expected to be bright 
because $F_g<F_e$. Indeed, a very weak bright resonance with contrast on the order of 0.3\% was observed 
when the laser power density was 0.14 mW/cm$^2$. However, as the power density of the exciting laser radiation was 
increased, the bright resonance disappeared and became dark for laser power densities greater than about 
0.6 mW/cm$^2$. At a laser power density of 2.9 mW/cm$^2$, the contrast of the dark resonance with respect to the intensity 
at the inflection point of the larger
structure was about 2\%. The resonances appeared to be somewhat narrower in the calculations than in the 
experiment. The reason for this discrepancy was most likely a residual transverse magnetic field on the order of a few 
milligauss. It should also be pointed out that the model made the simplifying assumption that the laser power density was 
constant over the laser beam width, whereas in reality there was a distribution of laser power densities.
Nevertheless, given the subtle nature of the effect, the agreement between experiment and theory was acceptable.

  The theoretical model unambiguously reproduced the change from a bright to a dark resonance for 
laser power densities above $I=0.4$ mW/cm$^2$. In order to understand the effect qualitatively, one must keep in mind 
that Doppler broadening allows the laser to excite the neighboring $F_g=2\longrightarrow F_e=2$ transition, and so the 
observed signal is the result of a superposition of the bright resonance at the $F_g=2\longrightarrow F_e=3$ transition 
and the dark resonance at the $F_g=2\longrightarrow F_e=2$ transition. The contrast of dark resonances increases 
significantly with power up to a certain point, while the contrast of bright resonances is less sensitive to power
(see, for example, ~\cite{Auzinsh:2008, Auzinsh:2009}). Therefore, at low power, the bright resonance contrast is 
stronger than the contrast of the distant dark resonance, but at higher power, the dark resonance drowns out the 
bright resonance.

\begin{figure*}[htbp]
	\centering
  \begin{tabular}{ccc}
		\resizebox{6.0cm}{!}{\includegraphics{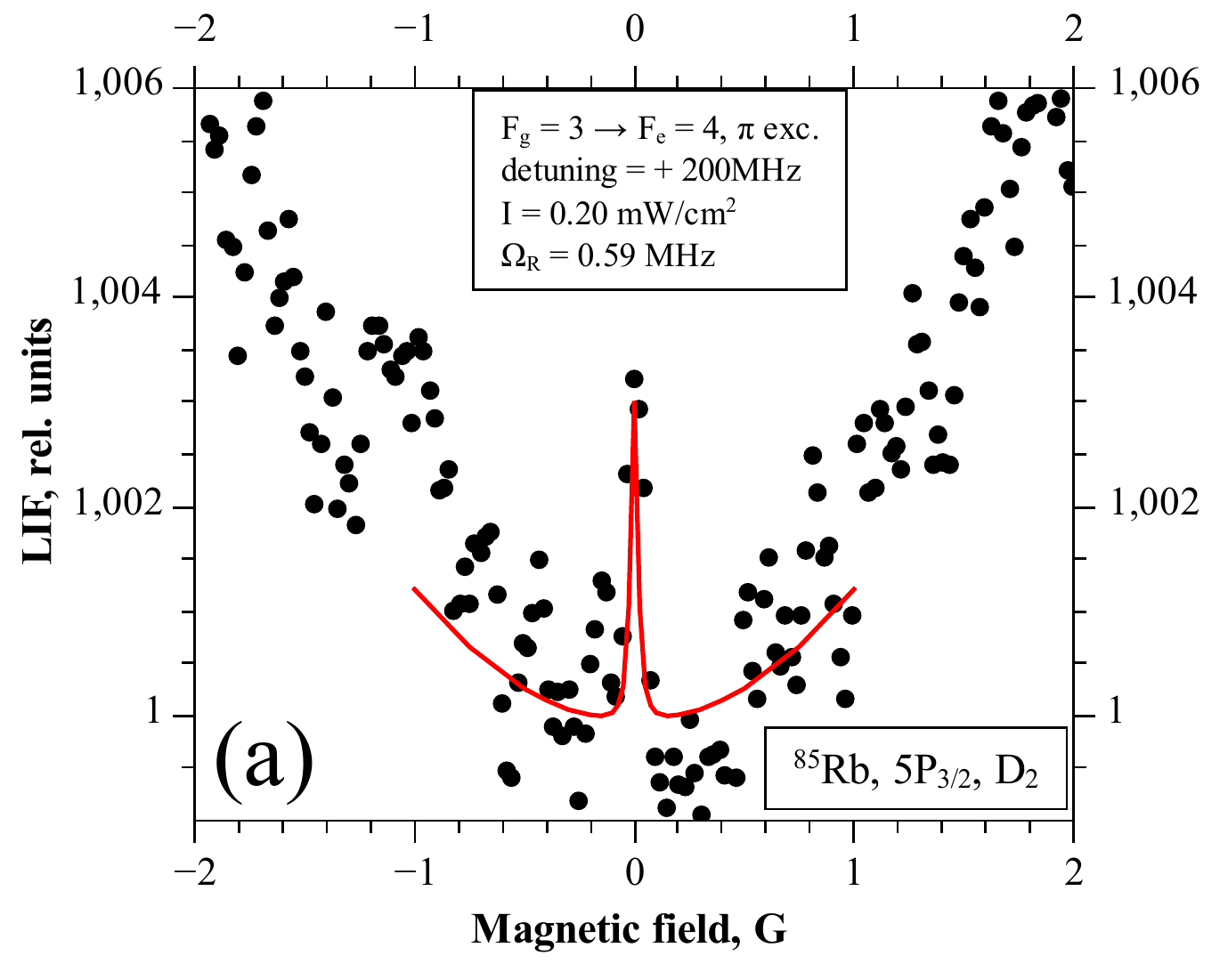}}&
		\resizebox{6.0cm}{!}{\includegraphics{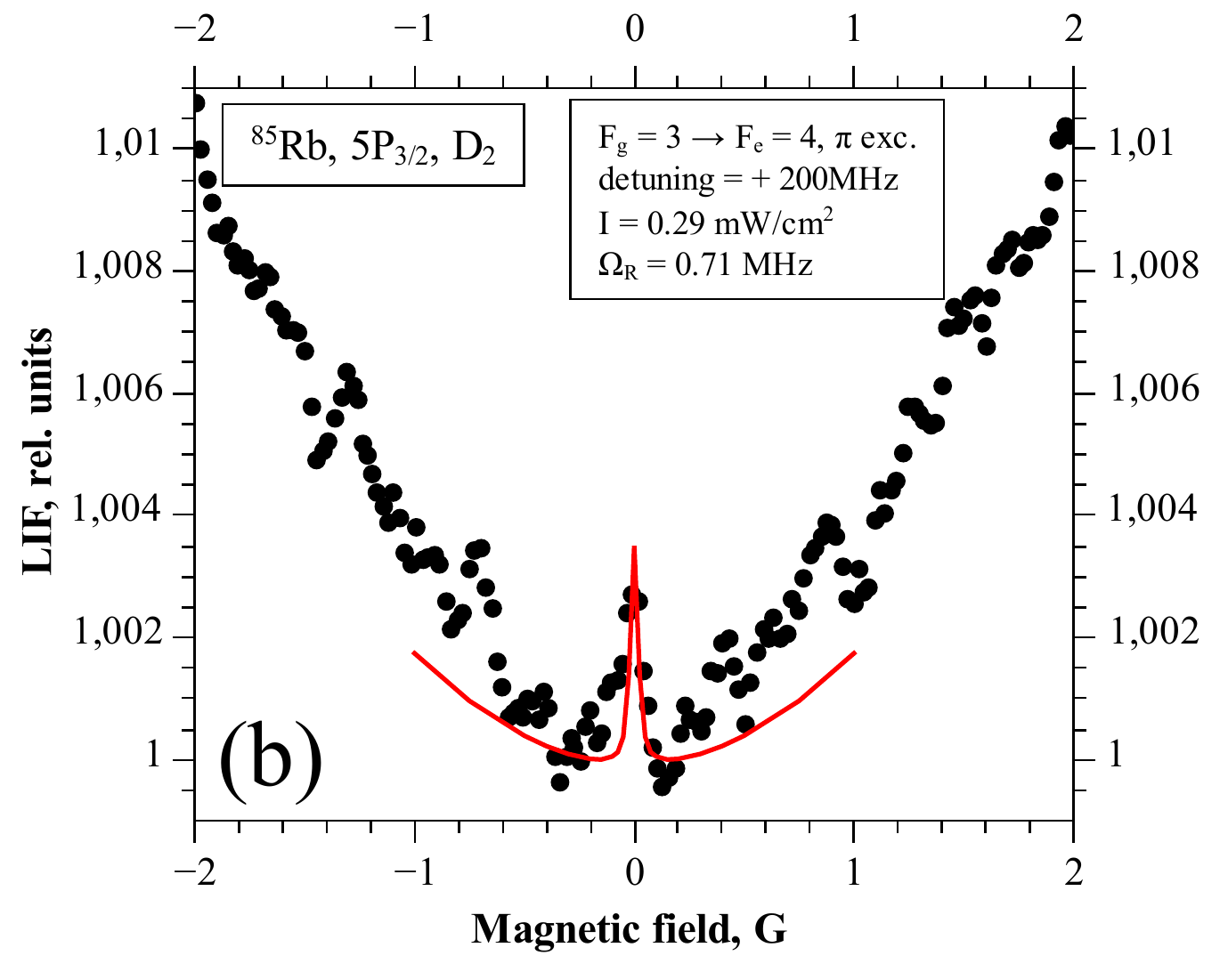}}&
		\resizebox{6.0cm}{!}{\includegraphics{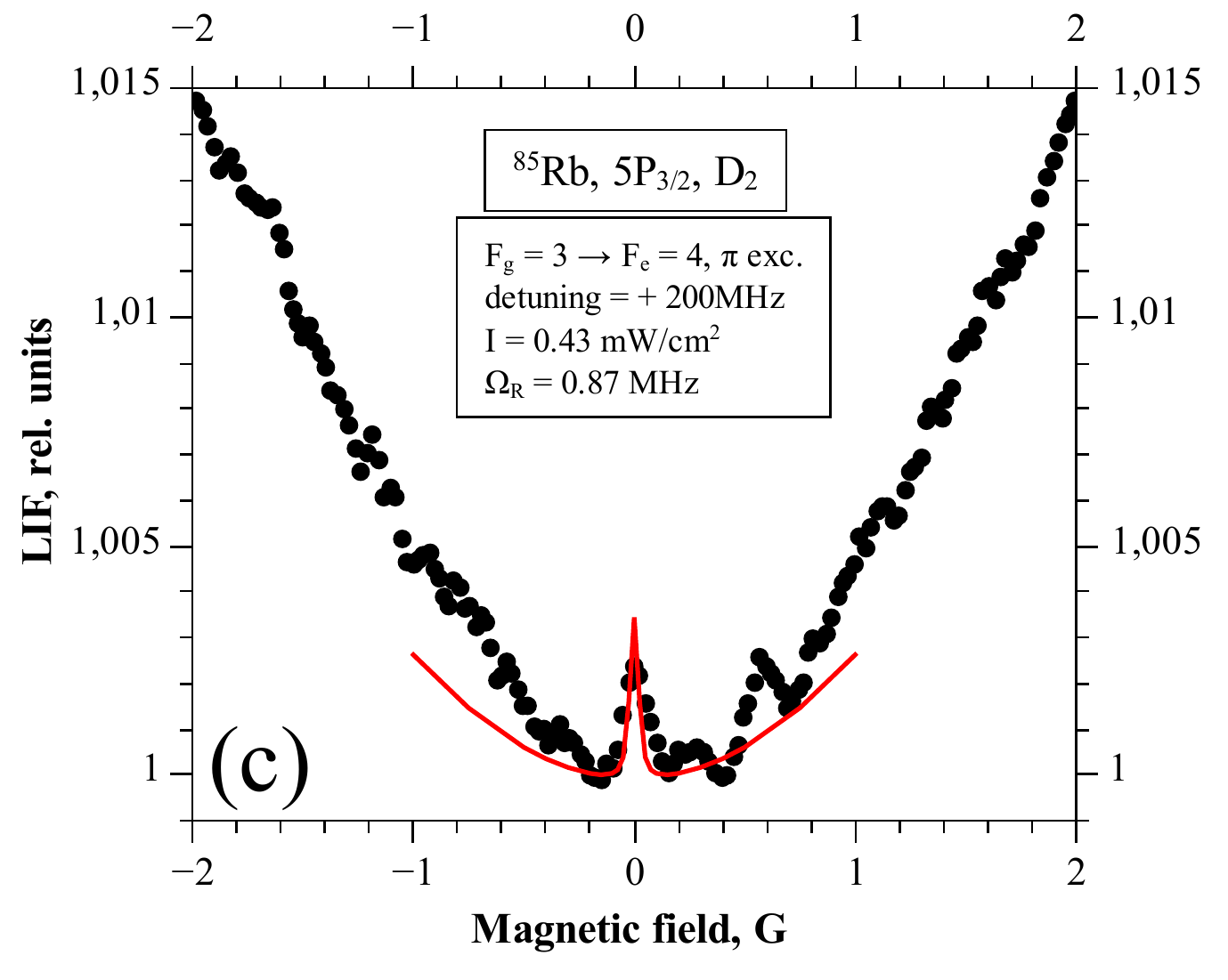}}\\
		\resizebox{6.0cm}{!}{\includegraphics{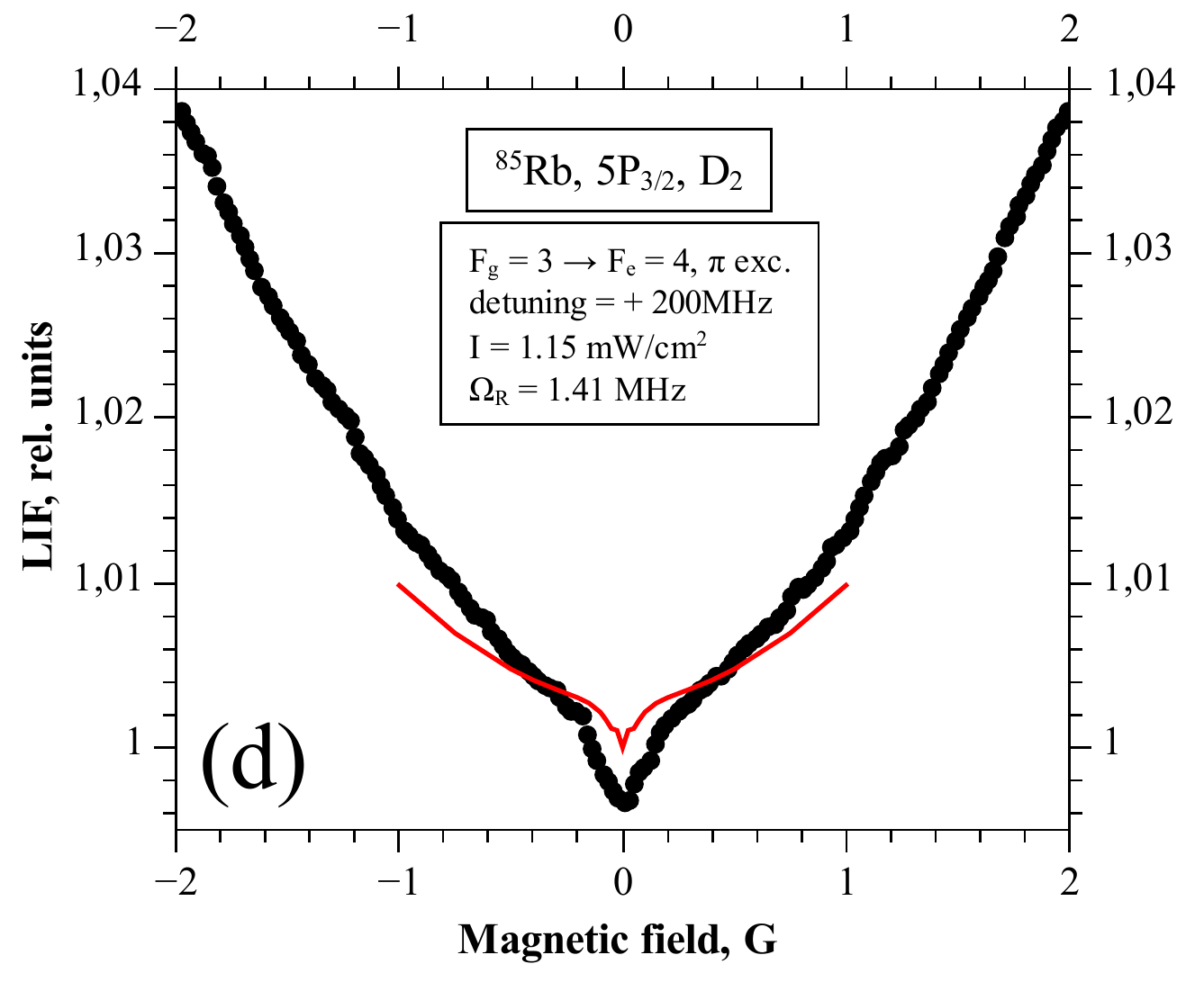}}&
		\resizebox{6.0cm}{!}{\includegraphics{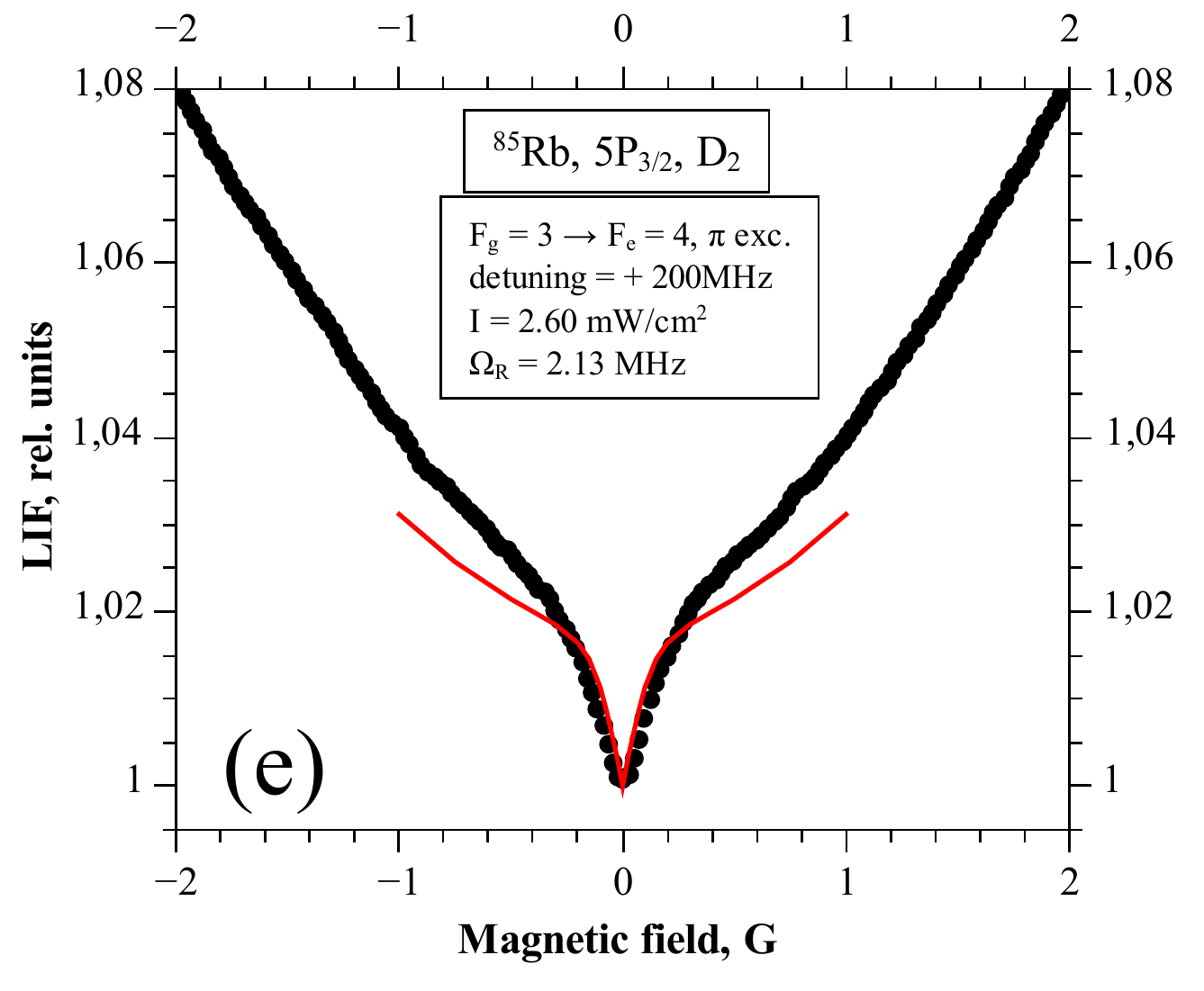}}&
		\resizebox{6.0cm}{!}{\includegraphics{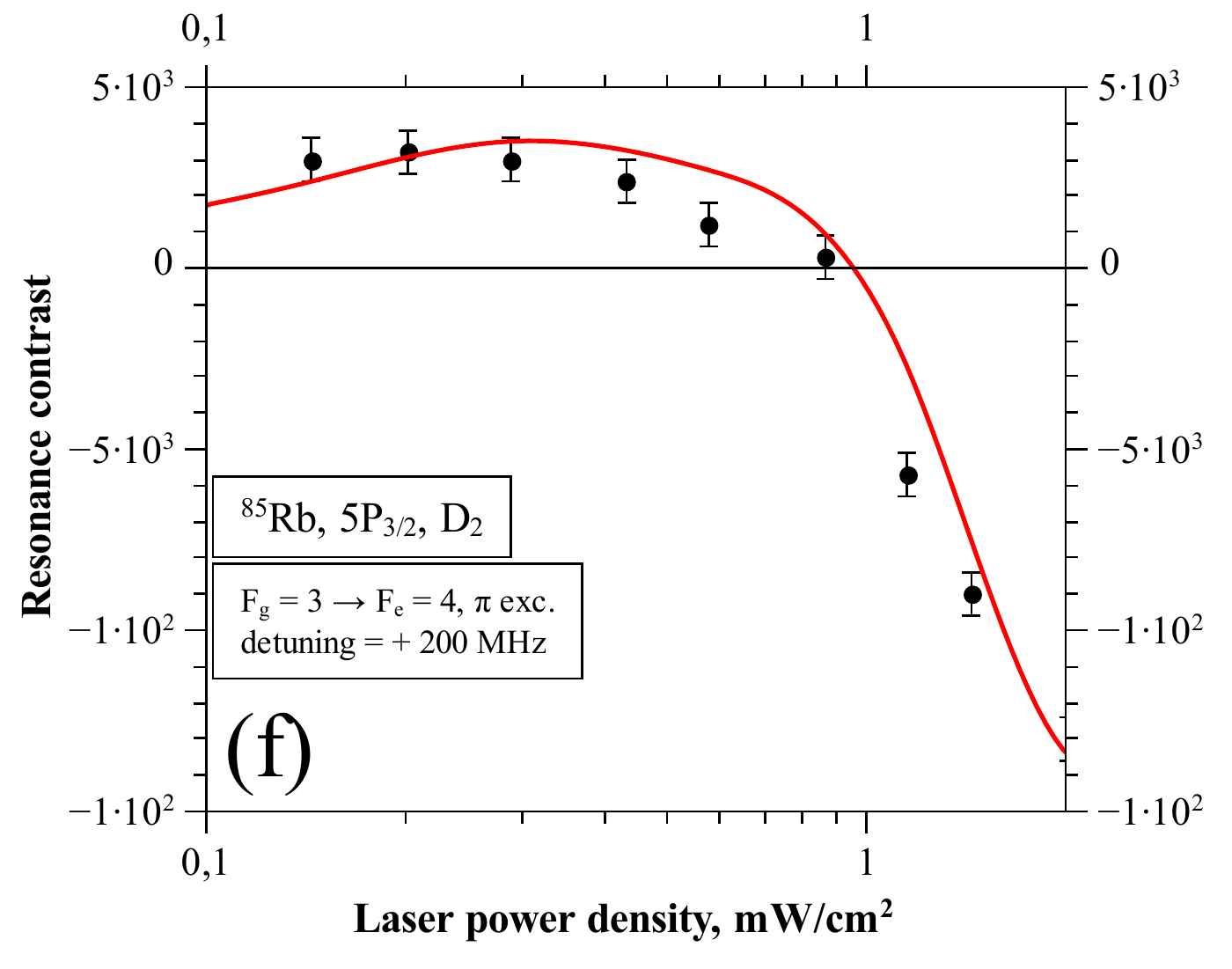}}
  \end{tabular}  
	\caption{\label{fig:rb85-34-det200} Bright and dark resonances 
obtained at various laser power densities near the $F_g=3\longrightarrow F_e=4$ transition of $^{85}$Rb. 
The laser was detuned from the exact transition by 200 MHz in the direction away from the 
$F_g=3\longrightarrow F_e=4$ transition for linearly polarized exciting laser radiation with a beam diameter of 
2.1 mm. 
Markers represent the experimental results, whereas the solid line represents the results of 
calculations. The final panel shows the contrast of the narrow resonance referenced to the estimated 
background due to the broad structure at zero magnetic field.  
}
\end{figure*}

Fig.~\ref{fig:rb85-34-det200} shows magneto-optical resonances obtained near the $F_g=3\longrightarrow F_e=4$ 
transition, but with the laser detuned by 200 MHz in the direction away from the $F_g=3\longrightarrow F_e=3$ 
transition. A bright resonance was observed at very low laser power densities. 
Similar to the case of the $F_g=2\longrightarrow F_e=3$ transition of $^{87}$Rb, the resonance 
became dark for laser power densities greater than 0.8 mW/cm$^2$. 
When the laser was tuned directly to the  $F_g=3\longrightarrow F_e=4$ transition of $^{85}$Rb, no bright resonance 
was observed even for laser power densities as low as 0.14 mW/cm$^2$. 
The reason was probably the influence of the strong dark resonance at the 
nearby $F_g=3\longrightarrow F_e=3$ transition, which could be excited for some velocity groups of atoms. Power broadening 
may also have played a small role in exciting this nearby transition. Even if the $F_g=3\longrightarrow F_e=3$ transition 
was excited only weakly, it could easily overwhelm the bright resonance at 
the $F_g=3\longrightarrow F_e=4$ transition.
The laser beam diameter for this series of measurements was 2.1 mm.

\begin{figure}[htbp]
	\centering
		\resizebox{\columnwidth}{!}{\includegraphics{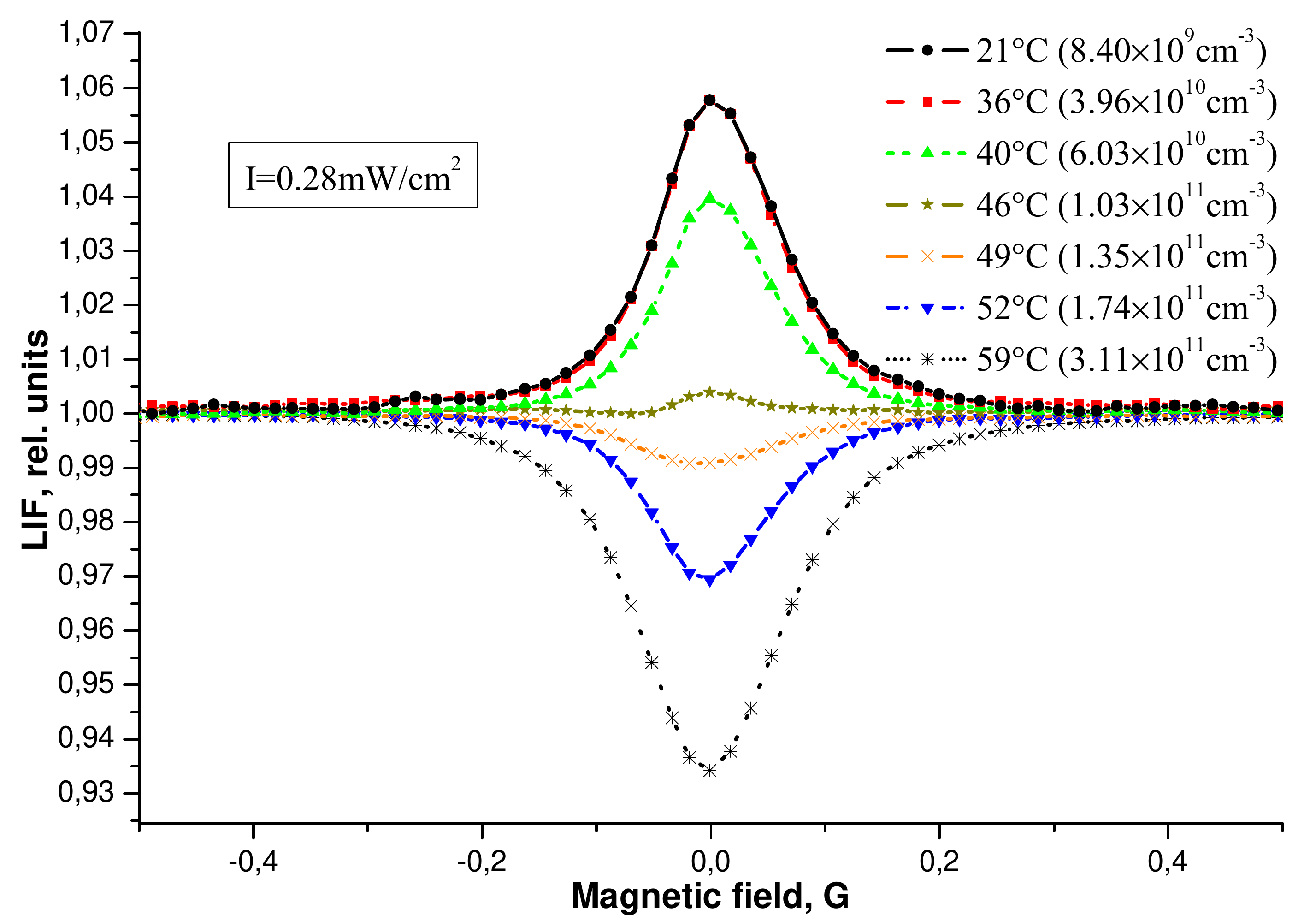}}
	\caption{\label{fig:temp_cirk} Dark and bright resonances 
obtained at various vapor temperatures at the $F_g=2\longrightarrow F_e=3$ transition of $^{87}$Rb for circularly 
polarized excitation at a laser power density of around 0.28 mW/cm$^2$. 
Experimental measurements are shown. The curves for 21$^o$C and 36$^o$C overlap.   
}
\end{figure}

\begin{figure}[htbp]
	\centering
		\resizebox{\columnwidth}{!}{\includegraphics{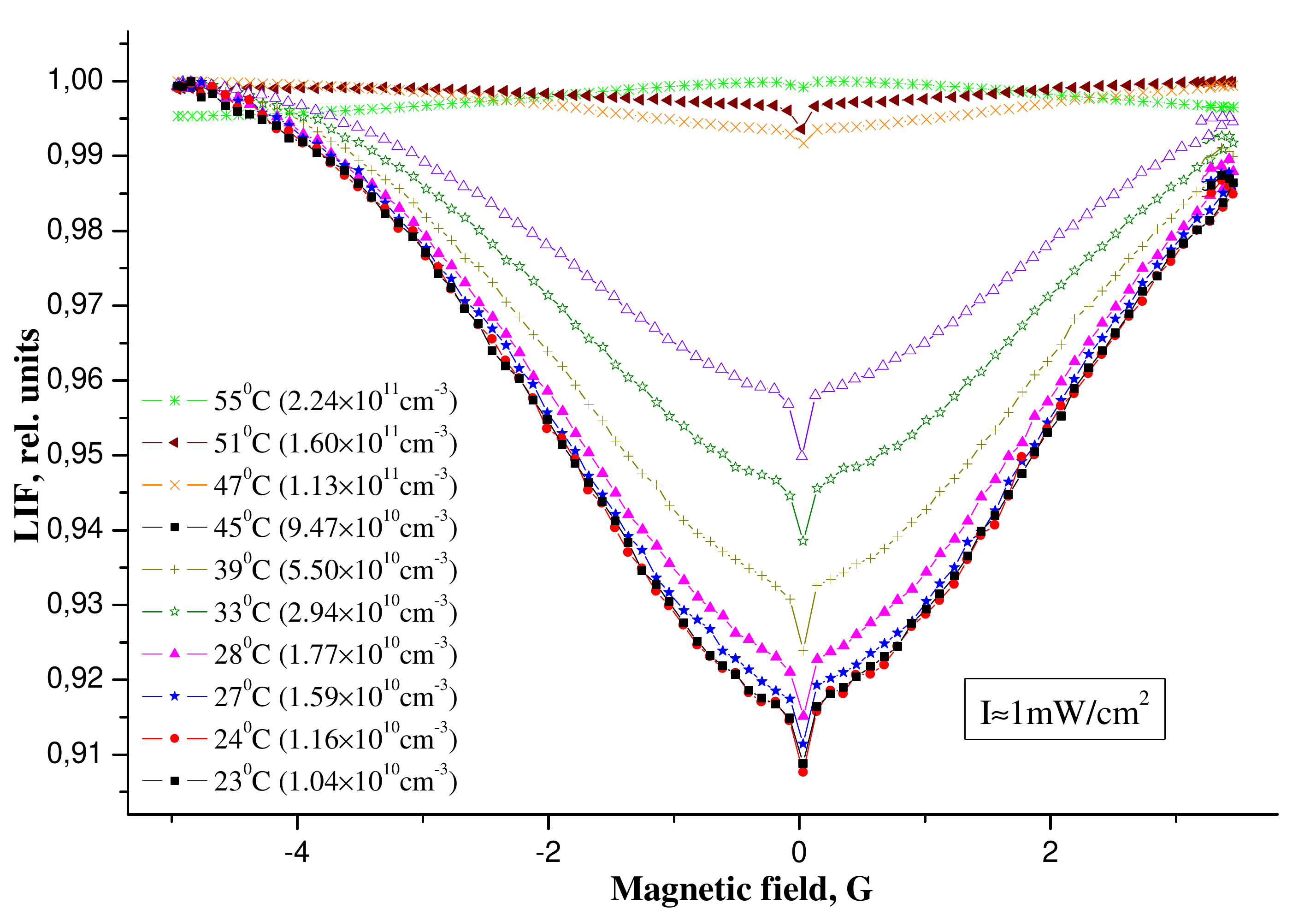}}
	\caption{\label{fig:temp_lin} Dark resonances 
obtained at various vapor temperatures at the $F_g=2\longrightarrow F_e=3$ transition of $^{87}$Rb for linearly 
polarized excitation at a laser power density of around 1 mW/cm$^2$. 
Experimental measurements are shown.    
}
\end{figure}

  Figure~\ref{fig:temp_cirk} shows the fluorescence signals obtained at the $F_g=2\longrightarrow F_e=3$ 
transition of $^{87}$Rb at various vapor temperatures and for circularly polarized excitation. 
The elevated temperatures were achieved by conducting hot air
around the cell as described in Section~\ref{Experiment:level1}.  As the temperature was 
increased, the contrast of the bright resonance decreased until the bright resonance disappeared at around $40^o$C. 
At higher
temperatures, a dark resonance was observed, and its contrast grew with increasing temperature. The change from bright 
to dark resonance at high temperatures was probably related to reabsorption. With each reabsorption cycle, information 
about the original coherent atomic state is lost. Unfortunately, our theoretical model does not include reabsorption 
effects at this point. Similar measurements were made for linearly polarized light (see Fig.~\ref{fig:temp_lin}).
In that case, although the contrast of the dark resonance decreased with temperature, it never changed sign. 

Our theoretical model does not include effects that become important at elevated temperatures, such as reabsorption. 
Thus, the change from bright to dark of the resonance at the  $F_g=2\longrightarrow F_e=3$ transition of $^{87}$Rb excited 
by circularly polarized light could not be reproduced by our model. Nevertheless, the model could be checked against the 
results at room temperature.

\begin{figure*}[htbp]
	\centering
		\resizebox{\columnwidth}{!}{\includegraphics{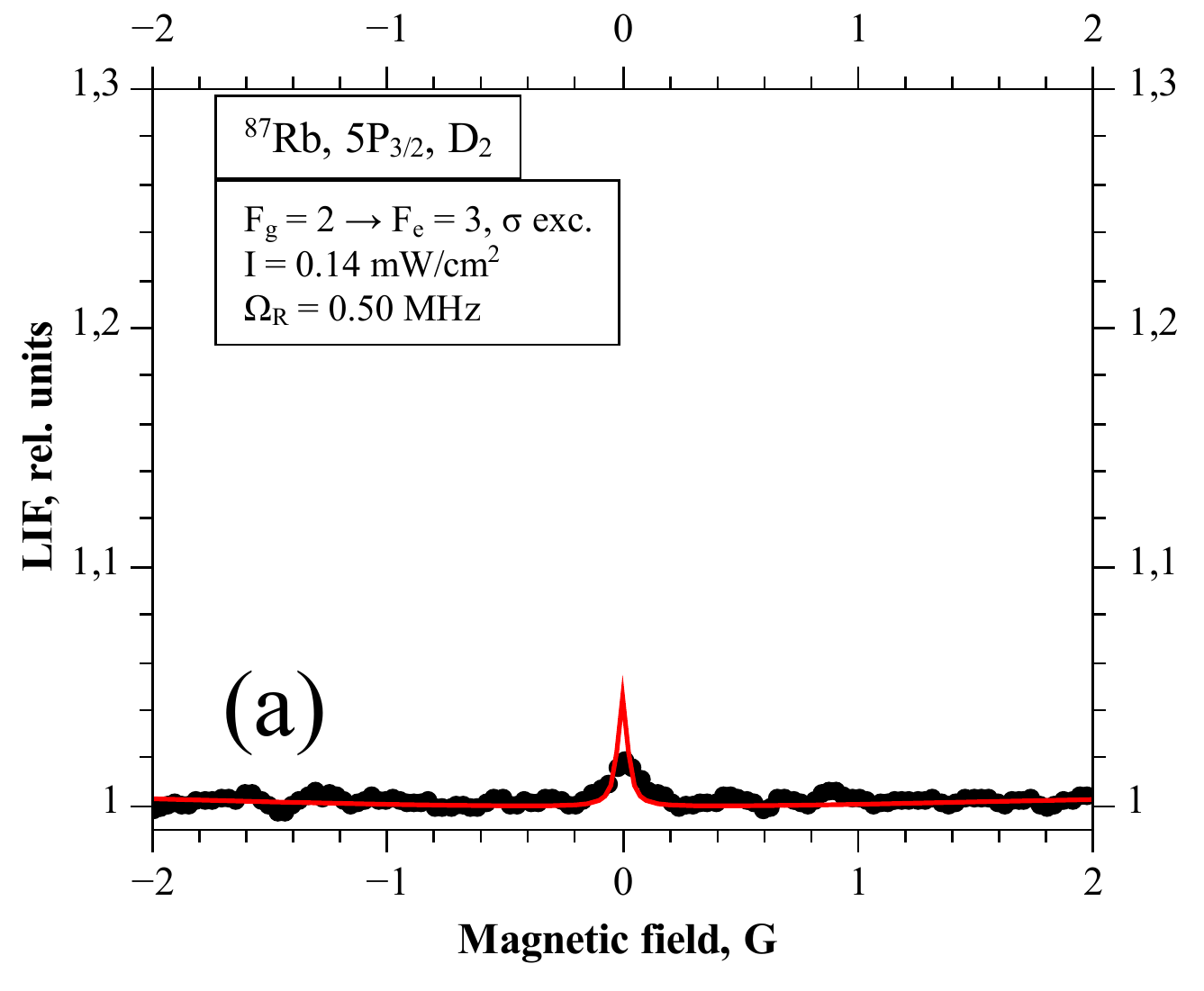}}
		\resizebox{\columnwidth}{!}{\includegraphics{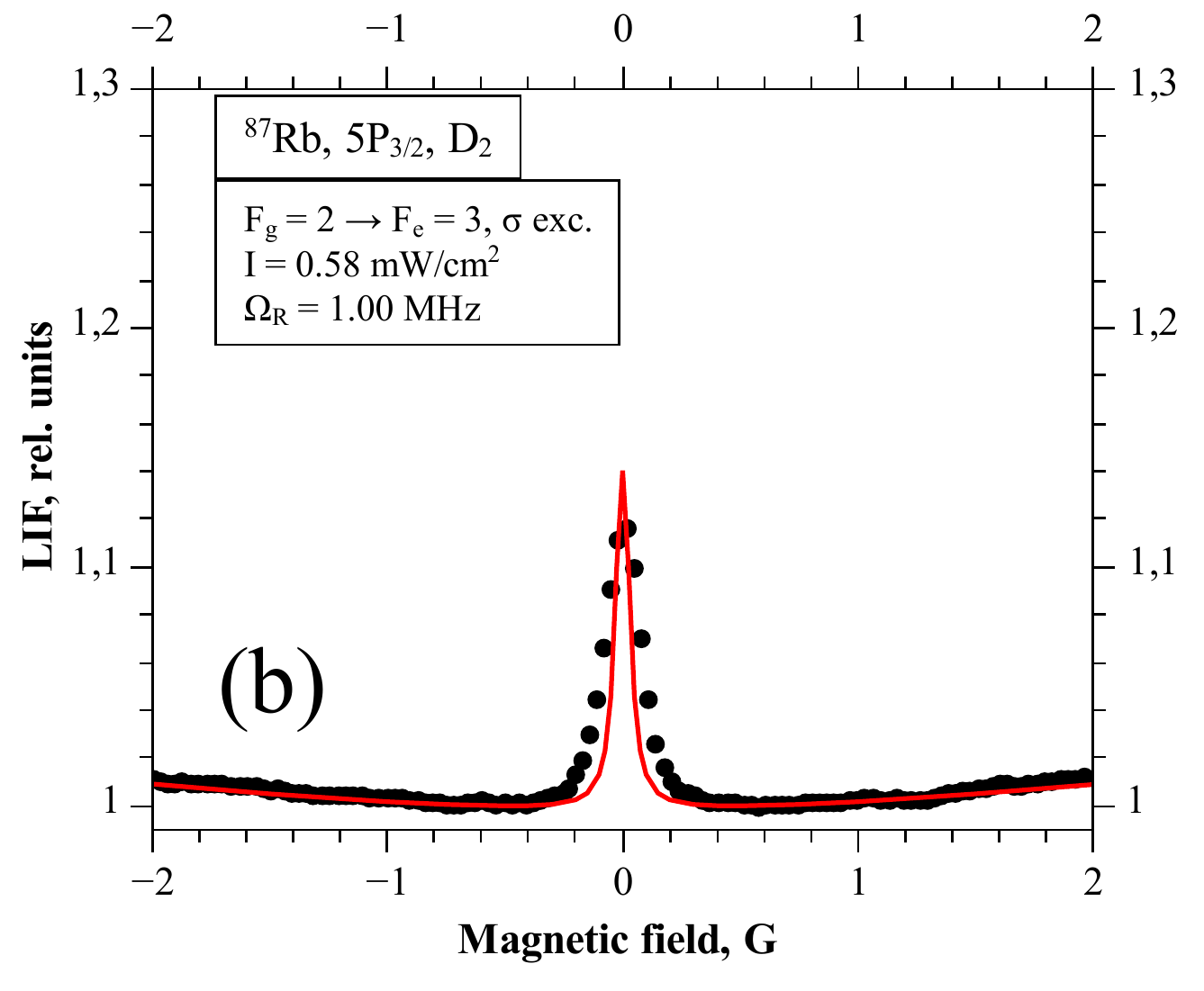}}
		\resizebox{\columnwidth}{!}{\includegraphics{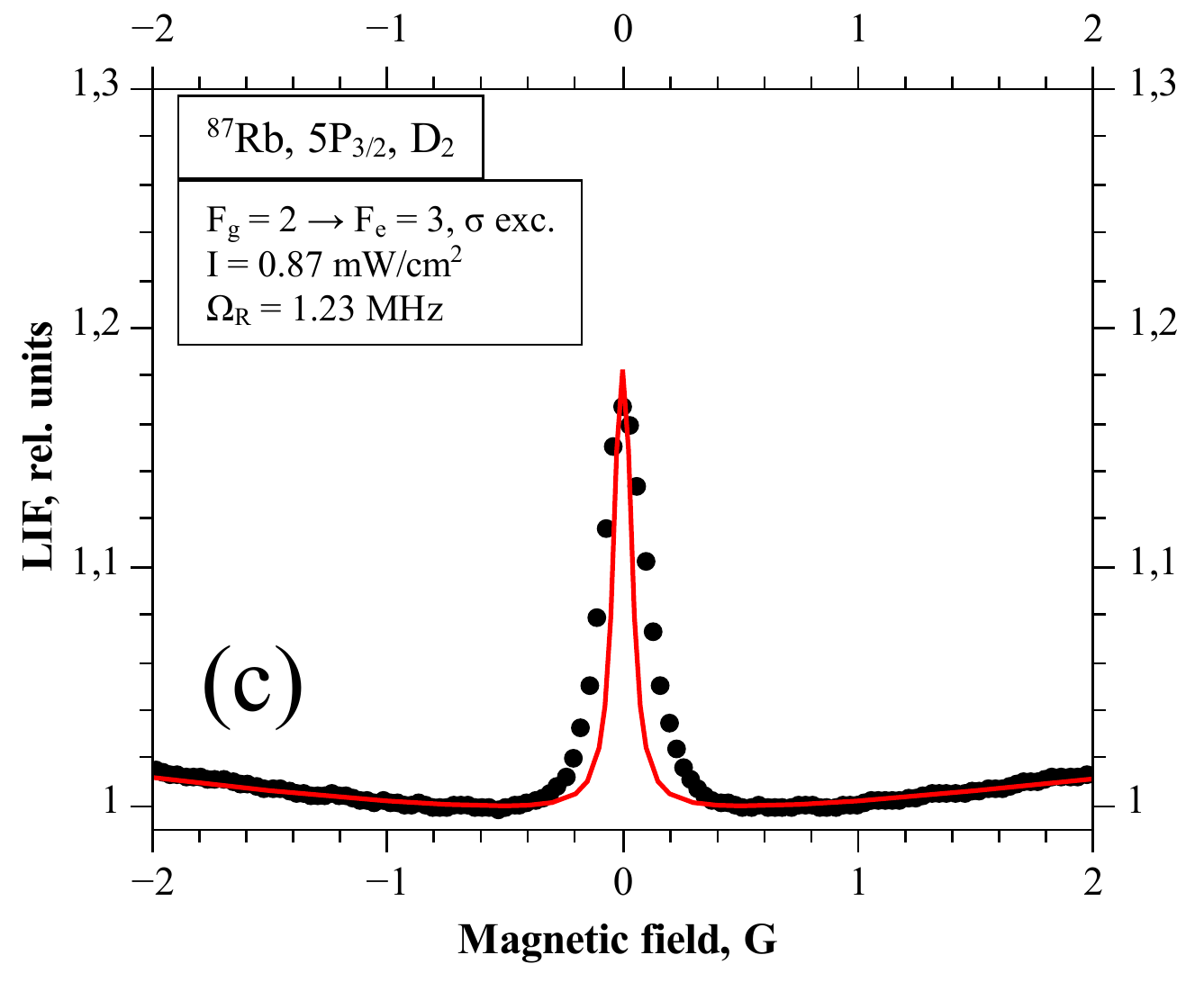}}
		\resizebox{\columnwidth}{!}{\includegraphics{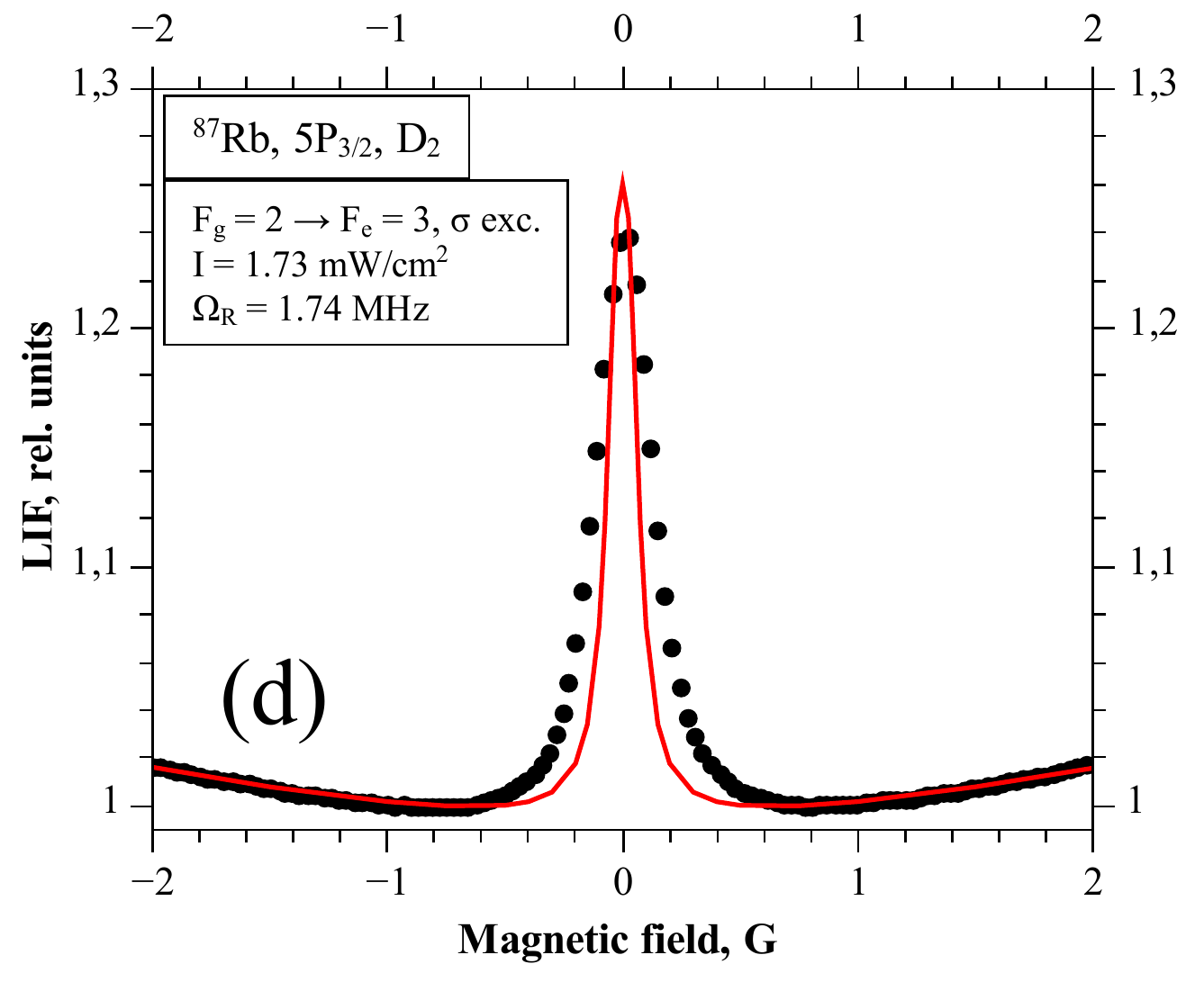}}
	\caption{\label{fig:rb87-circ} Bright resonances 
obtained at various laser power densities for circularly polarized excitation of 
the $F_g=2\longrightarrow F_e=3$ transition of $^{87}$Rb. 
Markers represent the experimental results, whereas the solid line represents the results of 
calculations. 
}
\end{figure*}

Fig.~\ref{fig:rb87-circ} shows measured and calculated signals for bright resonances obtained by exciting
the $F_g=2\longrightarrow F_e=3$ transition of $^{87}$Rb with circularly polarized laser radiation 
at various laser power densities. The rubidium vapor was at room temperature. The change in temperature is accompanied by 
a change from bright to dark resonance with contrasts on the order of six percent, much larger than in the case of linearly 
polarized excitation. 
The theoretical calculations show 
somewhat narrower resonances than the experiment. However, the large contrast is described well, and even the agreement 
between experiment and theory for the broad structure is better than in the case of the small bright resonances 
observed under linear excitation. 
Again, the model's idealization of the laser beam profile should be kept in mind.

\section{\label{Conclusion:level1}Conclusion}
  Nonlinear magneto-optical resonances in the hyperfine transitions of the $D_2$ line of atomic rubidium have been studied
under excitation by a single laser field.
When the exciting laser radiation was linearly polarized, bright resonances were observed 
at the $F_g=2 \longrightarrow F_e=3$ transition of $^{87}$Rb and at the 
$F_g=3 \longrightarrow F_e=4$ transition of $^{85}$Rb when the laser power density was very low. 
However, as the laser power density was 
increased above 0.6 mW/cm$^2$ or 0.8 mW/cm$^2$, respectively, these bright resonances became dark. 
The effect was described by a theoretical model based on the optical Bloch equations, 
which took into account all nearby hyperfine transitions, the mixing of magnetic sublevels in the external magnetic field, 
the coherence properties of the laser radiation, and the Doppler broadening. The parameter values in the model 
that could not be measured precisely were adjusted once for 
the entire set of measurements. These parameters were the coefficient relating transit relaxation rate to the laser beam 
diameter, the coefficient relating laser power density to the squared Rabi frequency, and the laser linewidth. 

  In the case of circularly polarized excitation of the $F_g=2\longrightarrow F_e=3$ transition of $^{87}$Rb, a bright 
resonance was observed at room temperature, but this bright resonance became dark as the vapor temperature was 
increased. The change might be related to reabsorption effects. 

The ability of the model to describe subtle effects, such as the change from bright to dark 
resonances in a system whose hyperfine structure was not 
resolved under Doppler broadening confirmed that the model is an adequate tool for studying nonlinear magneto-optical 
resonances in spite of its simplifying assumptions about the transit relaxation time and the laser power density 
distribution. However, the ability to change a bright resonance into a dark resonance by increasing the temperature suggests 
that in the future reabsorption effects should be included in the model.

\begin{acknowledgments}
We thank Andrey Jarmola for helpful advice and discussions. 
We gratefully acknowledge support from the Latvian State Research Programme Grant No. 2010/10-4/VPP-2/1, 
ERAF project Nr. 2DP/2.1.1.1.0/10/APIA/VIAA/036, and ESF project Nr. 2009/\\0223/1DP/1.1.1.2.0./09/APIA/VIAA/008.

\end{acknowledgments}
\bibliography{rubidium}
\end{document}